\documentclass[]{interact}

\usepackage{epstopdf}
\usepackage[caption=false]{subfig}
\usepackage{cite}
\usepackage{amsmath,amssymb,amsfonts}
\usepackage{textcomp}

\usepackage{graphicx}
\usepackage{subfig}
\usepackage{epstopdf}
\usepackage{comment}
\usepackage{amsthm}

\usepackage[natbibapa,nodoi]{apacite}
\theoremstyle{plain}

\theoremstyle{mystl}
\newtheorem{rmk}{Remark}
\newtheorem{lem}{Lemma}
\newtheorem{thme}{Theorem}
\newtheorem{dfn}{Definition}

\newtheorem{ass}{Assumption}

\begin{document}


\title{Observer-based Event-triggered Boundary
Control of the One-phase Stefan Problem}

\author{
\name{Bhathiya Rathnayake\textsuperscript{a}\thanks{CONTACT Bhathiya Rathnayake. Email: brm222@ucsd.edu} and Mamadou Diagne\textsuperscript{b}}
\affil{\textsuperscript{a}Department of Electrical and Computer
Engineering, University of California San Diego, 9500 Gilman Dr, La
Jolla, CA 92093, USA;\\ \textsuperscript{b}Department of Mechanical and Aerospace Engineering, University of California San Diego, 9500 Gilman Dr, La Jolla, CA 92093, USA}
}

\maketitle

\begin{abstract}
This paper provides an observer-based event-triggered boundary control strategy for the one-phase Stefan problem using the position and velocity measurements of the moving interface. The infinite-dimensional backstepping approach is used to design the underlying observer and controller. For the event-triggered implementation of the continuous-time observer-based controller, a dynamic event triggering condition is proposed. The triggering condition determines the times at which the control input needs to be updated. In between events, the control input is applied in a \textit{Zero-Order-Hold} fashion. It is shown that the dwell-time between two triggering instances is uniformly bounded below excluding \textit{Zeno behavior}. Under the proposed event-triggered boundary control approach, the well-posedness of the closed-loop system along with certain model validity conditions is provided. Further, using Lyapunov approach, the global exponential convergence of the closed-loop system to the setpoint is proved. A simulation example is provided to illustrate the theoretical results.
\end{abstract}


\section{Introduction}
In recent decades, the study of Stefan-type moving boundary problems driven by parabolic equations has found a new momentum due to the expansion of its interests into thriving areas of research such as additive manufacturing \citep{petrus2017online,chen2020enthalpy}, cyrosurgical operations \citep{rabin1997combined},  modeling of tumor growth \citep{friedman1999analysis}, and information diffusion in social media \citep{lei2013free}. The mathematical formulation of the classical one-phase Stefan problem for a monocomponent two phase material involves a diffusion partial differential equation (PDE) in cascade  with an ordinary differential equation (ODE). The PDE describes the thermal expansion of one phase along its dynamic spatial domain whereas the ODE captures the dynamics of the moving interface between the two phases \citep{rubinstein1979stefan}.

 The control of the Stefan problem deals with the stabilization of the temperature profile and the moving interface to a desired setpoint. During the past decade, inspired by the seminal work \citep{dunbar2003motion} on boundary control of the Stefan problem, numerous works have made contributions to tackle this challenging moving boundary PDE control problem \citep{petrus2012enthalpy,chen2020enthalpy,petrus2017online,chen2019enthalpy,maidi2014boundary,koga2018control,koga2020delay,koga2018input,ecklebe2021toward}.  An enthalpy-based full-state feedback boundary controller is proposed in \citep{petrus2012enthalpy} to ensure the asymptotic convergence of the closed-loop system to the setpoint. Compensating for the effect of input hysteresis, the authors of \citep{chen2019enthalpy} and \citep{chen2020enthalpy} develop full-state feedback and output feedback designs for the control of the Stefan problem, respectively. Using a geometric control approach, \citet{maidi2014boundary} achieves exponential stability of the closed-loop system for the one-phase Stefan problem via Lyapunov analysis. In recent years, Koga and coauthors have addressed the control of the Stefan problem in both theoretical settings \citep{koga2018control,koga2020materials} and application settings \citep{koga2020stabilization,koga2020materials} using the infinite-dimensional backstepping control approach which has been instrumental in the control of a wide variety of PDEs \citep{krstic2008boundary}. For the one-phase Stefan problem, the pioneering contribution \citep{koga2018control} discusses a full-state feedback control design along with robustness guarantees to parameter uncertainties, an observer design, and the corresponding output feedback control design under both Dirichlet and Neumann boundary actuations via backstepping approach, ensuring the exponential stability of the closed-loop system in $H_1$-norm. 
 
 In \citep{koga2021towards}, the authors consider the \textit{Zero-Order-Hold (ZOH)} implementation of the full-state feedback continuous-time stabilizing controller introduced in \citep{koga2018control}, leading to an aperiodic sampled-data control approach for the one-phase Stefan problem. Aperiodic sampled-data control strategies, which rely on nonuniform sampling schedules, are quite appealing as they point towards efficient use of limited hardware, software, and communication resources. In a relatively recent survey paper \citep{Laurentiu17}, comprehensive and relevant insights are provided into aperiodic sampled-data controller design, as well as limitations and challenges in their practical implementation. Although nonuniform sampling schedules in aperiodic sampled-data control offer increased flexibility, designers still have to manually select a schedule that adheres to the maximum allowable sampling diameter. This selection remains independent of the closed-loop system state, rendering the decision-making process open-loop. Event-triggered control strategies, on the other hand, provide a systematic solution to this drawback by bringing feedback to the sampling process. An event-triggered system transmits the system's states/outputs to a controller/actuator when the freshness in the sample exceeds an appropriate threshold involving the current state of the closed-loop system \citep{heemels2012introduction}. Only at the event times is the feedback loop closed, and between successive event times, the control is executed in an open-loop fashion. There have been numerous contributions during the past decade introducing event-triggered control strategies to control for PDE systems \citep{espitia2020observer,katz2020boundary,espitia2021event,diagne2021event,rathnayake2021observer,rathnayake2022sampled,wang2021event,rathnayake2022event}, to name a few.  For 2$\times$2 linear hyperbolic systems, an output feedback event-triggered boundary control strategies relying on dynamic triggering conditions is proposed  in  \citep{espitia2020observer}. The authors of \citep{espitia2021event} propose a full-state feedback event-triggered boundary control approach for reaction-diffusion PDEs with Dirichlet boundary conditions using ISS properties and small gain arguments. Using dynamic event-triggering conditions, the works \citep{rathnayake2021observer} and \citep{rathnayake2022sampled} develop output feedback control strategies for a class of reaction-diffusion PDEs under anti-collocated and collocated boundary sensing and actuation, respectively. A full-state feedback event-triggered boundary control strategy for the one-phase Stefan problem is proposed in \citep{rathnayake2022event} using a static triggering condition. 

This paper considers the output feedback boundary control of the one-phase Stefan problem using the position and velocity measurements of the moving interface. We propose an observer-based event-triggered boundary control strategy using a dynamic triggering condition under which we show that the closed-loop system is free from Zeno phenomenon. To the best of our knowledge, this work is the first to present an observer-based event-triggered boundary control approach for moving boundary type problems. In \citep{koga2020energy}, the authors propose a sampled-data observer-based boundary control design for the one-phase Stefan problem, yet with no theoretical guarantees. At event-times dictated by the proposed triggering condition, the continuous-time observer-based boundary control law derived in  \citep{koga2018control} is computed and applied to the plant in a ZOH fashion. The dynamic event-trigger makes use of a dynamic variable that depends on some information of the current states of the closed-loop system and the actuation deviation between the continuous-time boundary feedback and the event-triggered boundary control. We also prove that the closed-loop system is well-posed satisfying certain model validity conditions and globally exponentially converges to the setpoint subject to the proposed event-triggered control. The present work differs from \citep{rathnayake2021observer,rathnayake2022sampled} in that this paper involves a moving boundary making the Lyapunov analysis substantially different. Moreover, the Lyapunov candidate function involves the $H_1$-norm of the observer error target system unlike in \citep{rathnayake2021observer,rathnayake2022sampled} where the $L_2$-norm is sufficient. As opposed to \citep{rathnayake2021observer,rathnayake2022sampled}, dwell-times  between consecutive events in the Stefan problem has to be upper-bounded to maintain the positivity of the control input. Thus, careful design of the event-triggering mechanism is required to ensure that the minimal dwell-time is smaller than the largest dwell-time, otherwise, the well-posedness of the closed-loop system fails to exist.

The paper is organized as follows. Section 2 describes the one-phase Stefan problem and Section 3 presents the continuous-time observer-based backstepping boundary control and its emulation. In section 4, we introduce the event-triggered boundary control approach and present the main results of the paper. We conduct simulations in Section 5 and conclude the paper in Section 6. 

\textit{Notation:} $\mathbb{R}_+$ is the nonnegative real line whereas $\mathbb{N}$ is the set of natural numbers including zero. $t^{+}$ and $t^{-}$ respectively denote the right and left limit at time $t$. Let $u:[0,s(t)]\times \mathbb{R}_+\rightarrow\mathbb{R}$ be given. $u[t]$ denotes the profile of $u$ at certain $t\geq 0$, \textit{i.e.,} $\big (u[t]\big )(x)$, for all $x\in[0,s(t)]$. By $\Vert u[t]\Vert=\Big(\int_{0}^{s(t)}u^2(x,t)dx\Big)^{1/2}$ we denote $L_2(0,s(t))$-norm. $I_{m}(\cdot), $ and $J_{m} (\cdot)$ with $m$ being an integer respectively denote modified Bessel and (nonmodified) Bessel functions of the first kind.

\section{Description of the One-phase Stefan Problem}
\begin{figure}
\centering
\includegraphics[scale=0.8]{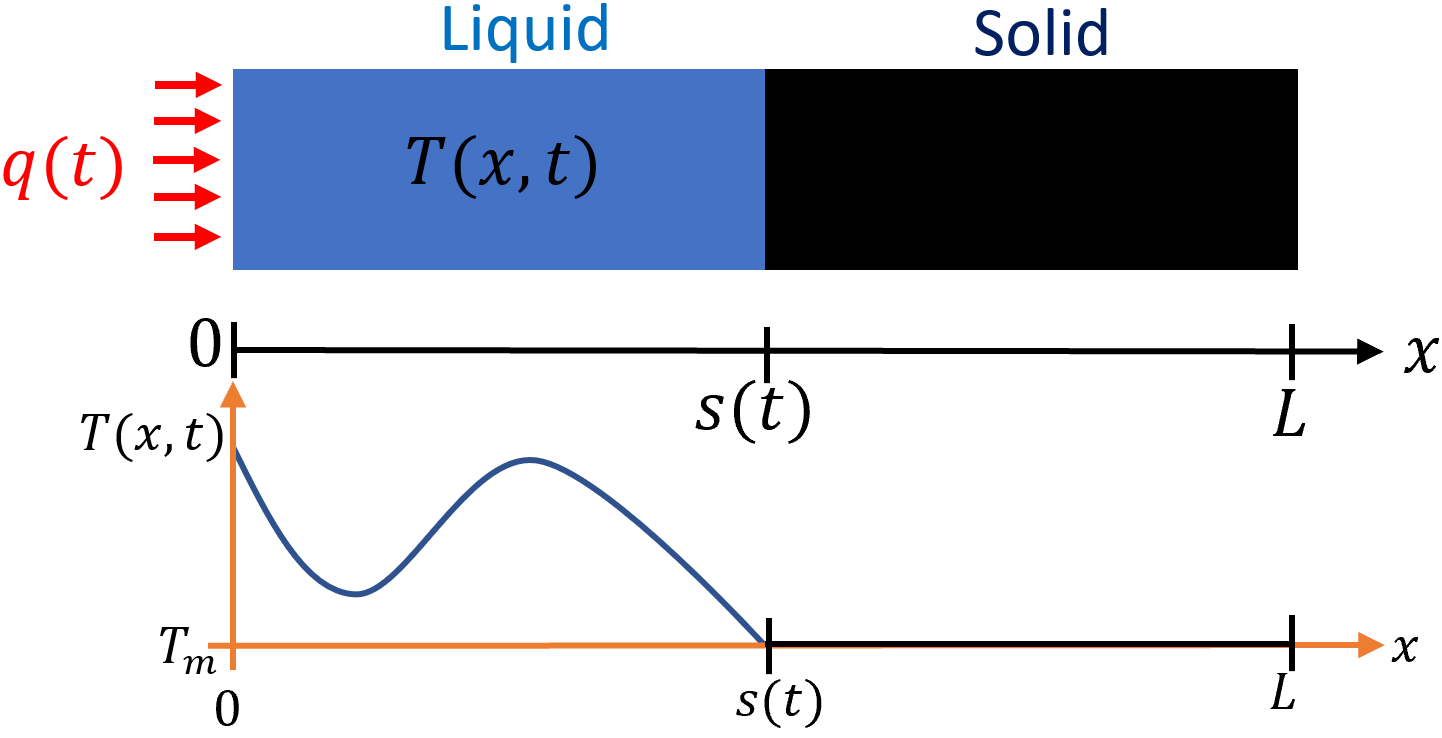}
\caption{Description of the one-phase Stefan problem.}
\label{fig1ww}
\end{figure}

Let us consider a physical model that describes the melting or
solidification process in a pure one-component material
of length $L$ in one dimension. The position $s(t)$ at which the phase transition occurs divides the domain $[0,L]$ into two time-varying sub-domains; the interval $[0, s(t)]$ containing the liquid phase,
and the interval $[s(t),L]$ containing the solid phase. The dynamics of the position of the liquid-solid interface is driven by a heat flux entering through the boundary at $x=0$ (the fixed boundary of the liquid phase). The heat equation coupled with the dynamics that describes the moving boundary is used to characterize the heat propagation in the liquid phase and the phase transition. Fig. \ref{fig1ww} illustrates this configuration.

Under the assumption that the temperature in the liquid phase is not lower than the melting temperature $T_{m}$ of the material, the conservation of energy and heat conduction laws can be used to derive the following PDE-ODE cascade system known as the one-phase Stefan Problem.
\begin{equation}\label{he1}
T_t(x,t)=\alpha T_{xx}(x,t), \text{   }\alpha:=\frac{k}{\rho C_p}, \text{   }0< x< s(t),
\end{equation}with the boundary conditions
\begin{align}\label{he3}
T(s(t),t)&=T_m,\\\label{he2}
-kT_x(0,t)&=q(t),
\end{align}
and the initial values
\begin{equation}\label{he4}
T(x,0)=T_0(x),\text{   }s(0)=s_0,
\end{equation}
where $T(x,t),q(t),\rho,C_p,$ and $k$ are the liquid phase distributed temperature, applied heat flux, the liquid density, the liquid heat capacity, and the liquid heat conductivity, respectively. By considering the local energy balance at the liquid-solid interface $x=s(t)$, the following ODE associated with the time-evolution of the spatial domain can be obtained: 
\begin{equation}\label{he5}
\dot{s}(t)=-\beta T_x(s(t),t),\text{   }\beta:=\frac{k}{\rho\Delta H^*},
\end{equation}
where $\Delta H^*$ is the latent heat of fusion. 

The validity of the physical model \eqref{he1}-\eqref{he5} relies on two physical conditions \citep{koga2019sampled}:
\begin{equation}\label{mv1}
T(x,t)\geq T_m,\text{   }\forall x\in[0,s(t)],\text{   }\forall t>0,
\end{equation}
\begin{equation}\label{mv2}
0<s(t)<L,\text{   }\forall t> 0.
\end{equation}
The first condition implies that the liquid phase should not be frozen to the solid phase from  the boundary $x=0$. The second condition implies that the material should not be completely melted or frozen to single phase through the disappearance of the other phase.

To be consistent with the conditions \eqref{mv1} and \eqref{mv2}, we make the following assumptions on the initial data: 
\medskip
\begin{ass}\label{ass1}
$s_0\in(0,L),T_0(x)\geq T_m$ for all $x\in[0,s_0],$ and $T_0(x)$ is continuously differentiable in $x\in[0,s_0]$. 
\end{ass}
\medskip

The well-posedness of the solution of the one-phase Stefan problem \eqref{he1}-\eqref{he5} has been presented in \citep{cannon1971remarks} and Lemma 1 in \citep{koga2019sampled} which we state as follows: 
\medskip
\begin{lem}\label{lem1}Subject to Assumption \ref{ass1}, if $q(t)$ is a bounded piece-wise continuous function producing nonnegative heat for a time interval, \textit{i.e.,} $q(t)\geq 0$, for all $t\in[0,\bar{t}]$, then there exists a unique solution for the Stefan problem \eqref{he1}-\eqref{he5} for all $t\in[0,\bar{t}],$ and the condition \eqref{mv1} is satisfied for all $t\in[0,\bar{t}].$ Furthermore, it holds that
\begin{equation}\label{dots}
\dot{s}(t)\geq 0,\text{   }\forall t\in[0,\bar{t}]. 
\end{equation}
\end{lem}

\section{Observer-based Backstepping Boundary Control and Emulation}
The steady-state solution $(T_{eq}(x),s_{eq})$ of the system \eqref{he1}-\eqref{he5} with zero input $q(t)=0$ delivers a uniform temperature distribution $T_{eq}(x)=T_m$ and a constant interface position determined by initial data. In \citep{koga2018control}, the authors proposed a continuous-time observer-based backstepping boundary controller using $s(t)$ and $T_x(s(t),t)$ (or equivalently $\dot{s}(t)$ due to \eqref{he5}) as the available measurements to exponentially stabilize the interface position $s(t)$ at a desired reference setpoint $s_{r}$ through the design of $q(t)$ as 
\begin{equation}\label{ctq}
    q(t)=-c\bigg(\frac{k}{\alpha}\int_{0}^{s(t)}\hat{u}(x,t)dx+\frac{k}{\beta}X(t)\bigg),
\end{equation}
where $c>0$ is the control gain and $\hat{u}(x,t)$ and $X(t)$ are reference error variables defined as
\begin{align}\label{u_hat}
    \hat{u}(x,t)&:=\hat{T}(x,t)-T_m,\\\label{xt}
    X(t)&:=s(t)-s_r.
\end{align}
Here, $\hat{T}(x,t)$ is the observer state which satisfies
\begin{align}\label{ob1}
    \hat{T}_t(x,t) = \alpha\hat{T}_{xx}(x,t)+p(x,s(t))\big(T_x(s(t),t)-\hat{T}_x(s(t),t)\big),
\end{align}
for $0<x<s(t)$ and
\begin{align}\label{ob3}
    \hat{T}(s(t),t)&=T_m,\\
    \label{ob2}
    -k\hat{T}_x(0,t)&=q(t),
\end{align}
with $p(x,s(t))$ being the observer gain given by

\begin{equation}\label{obg}
    p(x,s(t))=-\lambda s(t)\frac{I_1\Big(\sqrt{\frac{\lambda}{\alpha}\big(s^2(t)-x^2\big)}\Big)}{\sqrt{\frac{\lambda}{\alpha}\big(s^2(t)-x^2\big)}},\text{ }\lambda>0,
\end{equation}
for $0<x<s(t)$.  

We aim to stabilize the closed-loop system containing the plant \eqref{he1}-\eqref{he5} and the observer \eqref{ob1}-\eqref{obg} while sampling the continuous-time controller $q(t)$ given by \eqref{ctq} at a certain sequence of time instants $(t_{j})_{j\in\mathbb{N}}$. These time instants will be fully characterized later via a dynamic event-trigger. The control input is held constant between two consecutive time instants. Therefore, we define the control input for all $t\in[t_{j},t_{j+1}),j\in\mathbb{N}$ as
\begin{equation}\label{lln}
q_j=-c\Big(\frac{k}{\alpha}\int_{0}^{s(t_j)}\hat{u}(x,t_j)dx+\frac{k}{\beta}X(t_j)\Big).
\end{equation}
Accordingly, the boundary conditions \eqref{he2} and \eqref{ob2} are modified as follows:
\begin{equation}\label{she}
-kT_x(0,t)=q_j,
\end{equation}
\begin{equation}\label{she1}
-k\hat{T}_x(0,t)=q_j,
\end{equation}
for $t\in[t_j,t_{j+1}),j\in\mathbb{N}$.
Let the observer error state be defined as 
\begin{align}\label{u_tilde}
    \tilde{u}(x,t)&:=T(x,t)-\hat{T}(x,t).
\end{align}
Therefore, considering \eqref{he1},\eqref{he3},\eqref{ob1},\eqref{ob3},\eqref{she},\eqref{she1}, we can obtain that 
\begin{align}
    \label{obe1}
    \tilde{u}_t(x,t)&=\alpha\tilde{u}_{xx}(x,t)-p(x,s(t))\tilde{u}_x(s(t),t),\\
\label{obe2}
    \tilde{u}(s(t),t)&=0,\\\label{obe3}
    \tilde{u}_x(0,t)&=0,
\end{align}
for all $t>0$. In \citep{koga2018control}, the authors show that, subject to the following invertible backstepping transformation 
\begin{equation}\label{obeft}
    \tilde{u}(x,t)=\tilde{w}(x,t)+\int_{x}^{s(t)}P(x,y)\tilde{w}(y,t)dy,
\end{equation}
where 
\begin{equation}
    P(x,y)=\frac{\lambda}{\alpha}y\frac{I_1\Big(\sqrt{\frac{\lambda}{\alpha}(y^2-x^2)}\Big)}{\sqrt{\frac{\lambda}{\alpha}(y^2-x^2)}},\lambda>0,
\end{equation}
for $0\leq x\leq y\leq s(t)$, the observer error system \eqref{obe1}-\eqref{obe3} with the gain $p(x,s(t))$ chosen as in \eqref{obg} gets transformed into the following globally $H_1$-exponentially stable observer error target system
\begin{align}\label{tildw}
    \tilde{w}_t(x,t)&=\alpha\tilde{w}_{xx}(x,t)-\lambda\tilde{w}(x,t),\\
    \tilde{w}(s(t),t)&=0,\\\label{tildwq}
    \tilde{w}_x(0,t)&=0.
\end{align}
The inverse transformation of \eqref{obeft} is given by
\begin{equation}\label{inobst}
    \tilde{w}(x,t)=\tilde{u}(x,t)-\int_{x}^{s(t)}Q(x,y)\tilde{u}(y,t)dy,
\end{equation}
where 
\begin{equation}
    Q(x,y)=\frac{\lambda}{\alpha}y\frac{J_1\Big(\sqrt{\frac{\lambda}{\alpha}(y^2-x^2)}\Big)}{\sqrt{\frac{\lambda}{\alpha}(y^2-x^2)}},
\end{equation}
for $0\leq x\leq y\leq s(t)$. Considering \eqref{u_hat},\eqref{ob1},\eqref{ob3}, and \eqref{she1}, we can obtain that $\hat{u}$-system satisfies
\begin{align}\label{uhats1}
    \hat{u}_t(x,t)&=\alpha\hat{u}_{xx}(x,t)+p(x,s(t))\tilde{u}_x(s(t),t),\\\label{uhats2}
    \hat{u}(s(t),t)&=0,\\\label{uhats3}
    \hat{u}_x(0,t)&=-\frac{q_j}{k},
\end{align}
whereas considering \eqref{he5},\eqref{u_hat},\eqref{xt},\eqref{u_tilde}, we can show that the dynamics of $X(t)$ satisfies 
\begin{equation}\label{dum_x}
    \dot{X}(t)=-\beta\hat{u}_x(s(t),t)-\beta\tilde{u}_x(s(t),t).
\end{equation}

Let us consider the following backstepping transformation on $t\in[t_j,t_{j+1}),j\in\mathbb{N}$ as in \citep{koga2019sampled}, 
\begin{equation}\label{fdbt1}
\begin{split}
    \hat{w}(x,t)=&\hat{u}(x,t)-\frac{\beta}{\alpha}\int_{x}^{s(t)}\phi(x-y)\hat{u}(y,t)dy-\phi(x-s(t))X(t),
\end{split}
\end{equation}
where
\begin{equation}\label{fdbt2}
    \phi(x)=\frac{c}{\beta}x-\varepsilon,\text{ }\varepsilon>0. 
\end{equation}
It can be shown that the backstepping transformation \eqref{fdbt1},\eqref{fdbt2} transforms system \eqref{uhats1}-\eqref{dum_x} into the following system valid for $t\in[t_j,t_{j+1})$: 
\begin{align}\label{what1}
    \hat{w}_t(x,t)&=\alpha\hat{w}_{xx}(x,t)+\frac{c}{\beta}\dot{s}(t)X(t)+\tilde{w}_x(s(t),t)f(x,s(t)),\\
    \hat{w}(s(t),t)&=\varepsilon X(t),\\\label{what3}
    \hat{w}_x(0,t)&=-\frac{\varepsilon\beta}{\alpha}\hat{u}(0,t)+d(t),\\\label{whatx}
        \dot{X}(t)&=-cX(t)-\beta\hat{w}_x(s(t),t)-\beta\tilde{w}_x(s(t),t),
\end{align}
where
\begin{equation}\label{uuj}
\begin{split}
    f(x,s(t))=&p(x,s(t))-\frac{\beta}{\alpha}\int_{x}^{s(t)}\phi(x-y)p(y,s(t))dy+\beta\phi(x-s(t)),
\end{split}
\end{equation}
and
\begin{equation}\label{tnt}
\begin{split}
    d(t)=&\frac{c}{\alpha}\Big(\int_{0}^{s(t_j)}\hat{u}(y,t_j)dy-\int_{0}^{s(t)}\hat{u}(y,t)dy\Big)+\frac{c}{\beta}\big(X(t_j)-X(t)\big),
\end{split}
\end{equation}
for $t\in[t_{j},t_{j+1}),j\in\mathbb{N}$.

The inverse transform of \eqref{fdbt1} is given by 
\begin{equation}\label{intf}
\begin{split}
    \hat{u}(x,t)=&\hat{w}(x,t)-\frac{\beta}{\alpha}\int_{x}^{s(t)}\psi(x-y)\hat{w}(y,t)dy-\psi(x-s(t))X(t),
\end{split}
\end{equation}
for $t\in[t_j,t_{j+1}),j\in\mathbb{N}$ where
\begin{equation}\label{psi1}
    \psi(x)=e^{\nu x}(\zeta\sin(\omega x)+\varepsilon\cos(\omega x)),
\end{equation}
with 
\begin{equation}\label{sko}
    \nu=\frac{\beta\varepsilon}{2\alpha},\text{ }\omega=\sqrt{\frac{4\alpha c-(\varepsilon\beta)^2}{4\alpha^2}},\text{ }\zeta=-\frac{1}{2\alpha\beta\omega}\big(2\alpha c-(\varepsilon\beta)^2\big),
\end{equation}
and $0<\varepsilon<2\frac{\sqrt{\alpha c}}{\beta}$ to be chosen later.

\section{Observer-based Event-triggered Boundary Control}
\begin{figure}
\centering
\includegraphics[scale=0.8]{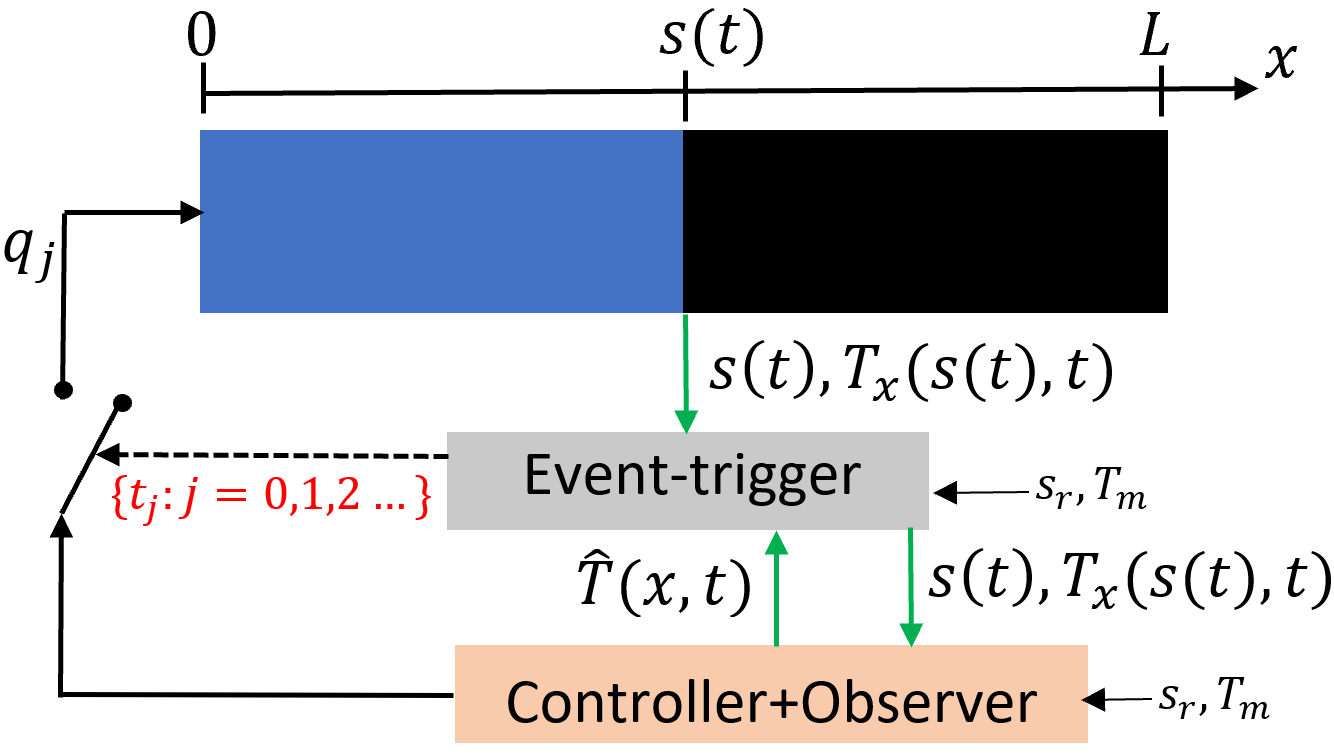}
\caption{Event-triggered control closed-loop system.}
\label{fig1wwr}
\end{figure}
Next we present the observer-based event-triggered boundary control approach for the one-phase Stefan problem. The closed-loop system consisting of the plant, the observer-based controller, and the event-trigger is shown in Fig. \ref{fig1wwr}.
\begin{dfn}\label{def1}Let $\eta,\gamma,\sigma,\mu_{1},\mu_{2},\mu_{3}>0$ be several design parameters. Further, let $c>0$ be the control gain in \eqref{lln}. The observer-based event-triggered boundary control strategy consists of two components:
\begin{enumerate}
\item The event-trigger: The set of event times $I=\big\{t_j\geq 0,j=0,1,2,\ldots\big\}$ is generated via the following rule with $t_0=0$:
\begin{equation}\label{eer}
    t_{j+1}=\min\big\{t_{j}+\frac{1}{c},\inf\big(\mathcal{S}(t,t_{j})\big)\big\}
\end{equation}
where \begin{equation}\label{bc2}
\begin{split}
&\mathcal{S}(t,t_j)=\big\{t\in\mathbb{R}_{+}|t>t_j\wedge\big(d^{2}(t)>\gamma m(t)\big)\big\}.
\end{split}
\end{equation}
Here $d(t)$ defined in \eqref{tnt}
for all for all $t\in[t_{j},t_{j+1})$ is the difference between the continuous time control input and the event-triggered control input, and $m(t)$ satisfies the ODE 
\begin{equation}\label{obetbc32}\begin{split}
\dot{m}(t)=&-\eta m(t)-\sigma d^{2}(t)+\mu_1\Vert \hat{u}[t]\Vert^2+\mu_2 X^2(t)+\mu_3\tilde{u}_x^2(s(t),t),\end{split}
\end{equation} for all $t\in(t_{j},t_{j+1})$ with $m(t_{0})=m(0)>0$ and $m(t_{j}^{-})=m(t_{j})=m(t_{j}^{+})$.

\item The control action: The boundary feedback control law is given by \eqref{lln}. 
\end{enumerate}
\end{dfn}
\begin{lem}\label{lemcon2}
    Along with Assumption 1, let us suppose the following Lipschitz continuity of $T_0(x)$ holds,
    \begin{equation}\label{ooly}
        0\leq T_0(x)-T_m\leq H(s_0-x),
    \end{equation}
    where $H$ is assumed to be known. For any initial temperature estimation $\hat{T}_0(x)$, any gain parameter of the observer $\lambda$, and any setpoint $s_r$, suppose that the following relations are satisfied respectively,
    \begin{align}\label{cn1}
        T_m+\hat{H}_\ell(s_0-x)&\leq\hat{T}_0(x)\leq T_m+\hat{H}_u(s_0-x),\\\label{cn2}
        \lambda&< \frac{4\alpha}{s_0^2}\frac{\hat{H}_\ell-H}{\hat{H}_u},\\\label{cn3}
        L>s_r&>s_0+\frac{\beta s_0^2}{2\alpha}\hat{H}_u,
    \end{align}
 where the parameters $\hat{H}_u$ and $\hat{H}_\ell$ satisfy $\hat{H}_u\geq \hat{H}_\ell>H$. Then, the event-triggered boundary control approach in Definition \ref{def1} generates positive heat, \textit{i.e.,} $q_j> 0$ for $j\in\mathbb{N}$ such that $j< j^*$ where \begin{equation}\label{J}j^*=\inf\big\{i\in\mathbb{N}\vert t_i= \sup(I)\big\},\end{equation} 
 with $I$ being the set of event times. Moreover, the closed-loop system containing the plant \eqref{he1},\eqref{he3},\eqref{he4},\eqref{he5},\eqref{lln},\eqref{she} and the observer \eqref{ob1},\eqref{ob3},\eqref{obg},\eqref{lln},\eqref{she1} has a unique solution satisfying the conditions \eqref{mv1}-\eqref{dots} for $t\in[0,\sup(I))$.
\end{lem}

\textit{Proof:} For $t\in (t_j,t_{j+1})$ where $j< j^*$, differentiating \eqref{ctq} along the solution of \eqref{uhats1}-\eqref{dum_x}, we can obtain that $$ \dot{q}(t)=-cq_j+ck\Big(1-\frac{1}{\alpha}\int_{0}^{s(t)}p(y,s(t))dy\Big)\tilde{u}_x(s(t),t).$$ Subject to the conditions \eqref{cn1} and \eqref{cn2}, following Lemma 3 and 4 in \citep{koga2018control} which make use of Maximum principle and Hopf's lemma, one can show that $\tilde{u}_x(s(t),t)>0$. Note from \eqref{obg} that $p(x,s(t))<0$ for $0<x<s(t)$. Thus, we can obtain that $\dot{q}(t)> -cq_j,$
for $t\in(t_j,t_{j+1})$ which we integrate in $t\in[t_j,t_{j+1})$ to obtain 
\begin{equation}\label{qt5}q(t)> \big(1-c(t-t_j)\big)q_j.\end{equation}
We use this recursively to derive that 
\begin{equation}
    q_j> q_0\prod_{i=0}^{j-1}\big(1-c(t_{i+1}-t_i)\big).
\end{equation}
One can easily show that $q_0>0$ when the conditions \eqref{cn1} and \eqref{cn3} are met. Furthermore, under the event-triggered boundary control approach in Definition \ref{def1}, it is ensured that $1-c(t_{i+1}-t_i)\geq 0$. Therefore, we have that $q_j> 0$ for all $j\in\mathbb{N}$ such that $j< j^*$. Thus, recalling Lemma \ref{lem1}, we can conclude that the plant \eqref{he1},\eqref{he3},\eqref{he4},\eqref{he5},\eqref{lln},\eqref{she} has a unique solution satisfying the conditions \eqref{mv1} and \eqref{dots} in the interval $[0,\sup(I))$. The observer error system \eqref{obe1}-\eqref{obe3} has a unique solution due to the transformation \eqref{obeft} and as the observer error target system \eqref{tildw}-\eqref{tildwq} admits a unique solution. Thus, the observer PDE \eqref{ob1},\eqref{ob3},\eqref{obg},\eqref{lln},\eqref{she1} also admits a unique solution in the interval $[0,\sup(I))$ due to \eqref{u_tilde}. Again, subject to the conditions \eqref{cn1} and \eqref{cn2}, one can show that $\tilde{u}(x,t)<0$ as in Lemma 3 and 4 in \citep{koga2018control}. Thus, considering \eqref{mv1},\eqref{u_hat},\eqref{u_tilde}, we can verify that $\hat{u}(x,t)>0$. Further, note from \eqref{qt5} that $q(t)> 0$ for $t\in[0,\sup(I))$. Thus, considering \eqref{ctq}, we can show that $X(t)< 0$ for $t\in[0,\sup(I))$  which in combination with \eqref{dots} and \eqref{xt} leads to the property \eqref{mv2} for $t\in[0,\sup(I))$. \hfill $\qed$
\begin{lem}\label{pom}Under the definition of the event-trigger \eqref{eer}-\eqref{obetbc32}, it holds that $d^2(t)\leq\gamma m(t)$ and $m(t)>0$, for all $t\in[0,\sup(I))$.
\end{lem}
\textit{Proof:} According to Definition \ref{def1}, the triggering of events guarantee that $d^{2}(t)\leq \gamma m(t),t\in [0,\sup(I))$. This inequality in combination with \eqref{obetbc32} yields:  
\begin{equation}\label{lem1e1}
\begin{split}
\dot{m}(t)\geq&-(\eta+\gamma\sigma) m(t)+\mu_{1}\Vert\hat{u}[t]\Vert^{2}+\mu_{2}X^2(t)+\mu_{3}\tilde{u}_x^2(s(t),t),
\end{split}
\end{equation}for $ t\in(t_{j},t_{j+1})$ and $j\in\mathbb{N}$ such that $j<j^*$ where $j^*$ is given by \eqref{J}. Thus, considering the time-continuity of $m(t)$, we can obtain the following estimate:
\begin{equation}\label{fa}
\begin{split}
&m(t)\geq m(t_{j})e^{-(\eta+\gamma\sigma)(t-t_{j})}+\int_{t_{j}}^{t}e^{-(\eta+\gamma\sigma)(t-\tau)}\big(\mu_{1}\Vert\hat{u}[\tau]\Vert^{2}+\mu_{2}X^2(t)+\mu_{3}\tilde{u}^2_x(s(\tau),\tau)\big)d\tau,
\end{split}
\end{equation}
for all $t\in[t_{j},t_{j+1}]$. From Definition \ref{def1}, we have that \mbox{$m(t_{0})=m(0)>0$}. Therefore, it follows from \eqref{fa} that $m(t)>0$  for all $t\in[0,t_{1}]$. Again using \eqref{fa} on $[t_{1},t_{2}]$, we can show that $m(t)> 0$  for all $t\in[t_{1},t_{2}]$. Applying the same reasoning successively to the future intervals, it can be shown that $m(t)> 0$ for $t\in [0,\sup(I))$.\hfill$\qed$

\begin{lem}\label{lem2} For $d(t)$ given by \eqref{tnt}, it holds that\begin{equation}\label{ghm}
\dot{d}^{2}(t)\leq \theta_{0} d^{2}(t)+\theta_{1}\Vert\hat{u}[t]\Vert^{2}+\theta_{2}X^2(t)+\theta_3\tilde{u}_x^2(s(t),t),
\end{equation}
where
\begin{align}\label{thetas}
    \theta_0= 4c^2,\text{ }
    \theta_1=\frac{4c^4 L}{\alpha^2},\text{ }
    \theta_2=\frac{4c^4}{\beta^2},\text{ }\theta_3=4c^2\Upsilon^2,
\end{align}
with $$\Upsilon = \max_{0\leq s(t)\leq L}\Big\{\Big\vert1-\frac{1}{\alpha}\int_{0}^{s(t)}p(y,s(t))dy\Big\vert\Big\}$$ for all $t\in(t_{j},t_{j+1})$ and $j\in\mathbb{N}$ such that $j<j^*$ where $j^*$ is given by \eqref{J}.\end{lem}

\textit{Proof:} For $t\in(t_j,t_{j+1}),j\in\mathbb{N}$, differentiating $d(t)$ given by \eqref{tnt} along the solution of \eqref{uhats1}-\eqref{dum_x} and using \eqref{tnt}, we can obtain that
\begin{equation}\label{ddot_dum}
\begin{split}
    \dot{d}(t)&=c\hat{u}_x(0,t)-\frac{c}{\alpha}\int_{0}^{s(t)}p(y,s(t))dy\tilde{u}_x(s(t),t)+c\tilde{u}_x(s(t),t)\\
    &=cd(t)+\frac{c^2}{\alpha}\int_{0}^{s(t)}\hat{u}(y,t)dy+\frac{c^2}{\beta}X(t)+c\Big(1-\frac{1}{\alpha}\int_{0}^{s(t)}p(y,s(t))dy\Big)\tilde{u}_x(s(t),t).
\end{split}
\end{equation}
Using Cauchy-Schwarz inequality and Young's inequality along with the fact that $0<x<s(t)<L$, we can obtain from \eqref{ddot_dum} that
\begin{equation}\label{hdum1}
    \begin{split}
        \dot{d}^2(t)\leq &4c^2 d^2(t)+\frac{4c^4L}{\alpha^2}\Vert\hat{u}[t]\Vert^2+\frac{4c^4}{\beta^2}X^2(t)+4c^2\Upsilon^2\tilde{u}_x^2(s(t),t),
    \end{split}
\end{equation}
for all $t\in(t_{j},t_{j+1})$ and $j\in\mathbb{N}$ such that $j<j^*$ where $j^*$ is given by \eqref{J}.
\hfill $\qed$
\medskip
\subsubsection{Avoidance of Zeno behavior}
\begin{thme}\label{ffg} Let $\delta\in (0,1)$ be chosen such that
\begin{equation}\label{delts_cn}
    \delta < \frac{1}{1+c},
\end{equation}
where $c$ is the controller gain in \eqref{lln}. Then, under the observer-based event-triggered boundary control in Definition \ref{def1}, with $\mu_{1},\mu_{2},\mu_{3}$ chosen as\begin{equation}\label{betas}
\mu_{1}=\frac{\theta_{1}}{\gamma(1-\delta)},\hspace{5pt}\mu_{2}=\frac{\theta_{2}}{\gamma(1-\delta)},\hspace{5pt}\mu_{3}=\frac{\theta_{3}}{\gamma(1-\delta)},
\end{equation}where $\theta_{1},\theta_{2},\theta_{3}$ given by \eqref{thetas},  there exists $\tau>0$ such that $t_{j+1}-t_j\geq \tau$ for all $j\in\mathbb{N}$ and any choice of admissible initial conditions $T_0(x),\hat{T}_0(x)$ and $s_0$, and desired liquid-solid interface $s_r$ which satisfy Assumption \ref{ass1} and the conditions \eqref{ooly}-\eqref{cn3}.
\end{thme}
\textit{Proof:}  Let us assume that an event has occurred at $t=t_j$ for some $j\in\mathbb{N}$ such that $j<j^*$ where $j^*$ is given by \eqref{J}. Furthermore, without loss of generality, let us assume that the set $\mathcal{S}(t,t_j)$ given by \eqref{bc2} is not empty (otherwise the next event time is $t_{j+1}=t_j+1/c$). Therefore, according to \eqref{eer}, we have that
\begin{equation}
    t_{j+1}=\inf\big(\mathcal{S}(t,t_j)\big).
\end{equation}
In this case, we have that $d^2(t)<\gamma m(t)$ for $t\in(t_j,t_{j+1})$ and $d^2(t_{j+1}^{-})=\gamma m(t_{j+1}^{-})$. Further, from Lemma \ref{pom}, we have that $m(t)>0$ for $t\in [0,\sup(I))$.

Let us define the function
\begin{equation}\label{phit}
    \Phi(t):=\frac{d^2(t)-\gamma(1-\delta)m(t)}{\gamma\delta m(t)}.
\end{equation}
Note that $\Phi(t)$ is continuous in $[t_{j},t_{j+1})$ and $\Phi(t_{j+1}^{-})=1$. A lower bound for the dwell-times is given by the time it takes for the function $\Phi$ to go from $\Phi(t_{j})$ to $\Phi(t_{j+1}^{-})=1,$ where $\Phi(t_{j})<0$, which holds since $d(t_{j})=0$. Therefore, by the intermediate value theorem, there exists a $\hat{t}_{j}>t_{j}$ such that $\Phi(\hat{t}_{j})=0$ and $\Phi(t)\in[0,1]$ for $t\in[\hat{t}_{j},t_{j+1}^-]$. The time derivative of $\Phi$ on $[\hat{t}_{j},t_{j+1})$ is given by
\begin{equation}
    \dot{\Phi}(t)=\frac{2d(t)\dot{d}(t)-\gamma(1-\delta)\dot{m}(t)}{\gamma\delta m(t)}-\frac{\dot{m}(t)}{m(t)}\Phi(t).
\end{equation}
From Young's inequality, we have that
\begin{equation}
    \dot{\Phi}(t)\leq \frac{d^2(t)+\dot{d}^2(t)-\gamma(1-\delta)\dot{m}(t)}{\gamma\delta  m(t)}-\frac{\dot{m}(t)}{m(t)}\Phi(t).
\end{equation}
Using Lemma \ref{lem2} and \eqref{obetbc32}, we can show that
\begin{equation}\label{ggdum}
    \begin{split}
        \dot{\Phi}(t)\leq& \frac{(1+\theta_0+\gamma(1-\delta)\sigma) d^2(t)}{\gamma\delta m(t)}+\frac{1-\delta}{\delta}\eta+\frac{\gamma \delta\sigma d^2(t)}{\gamma\delta m(t)}\Phi(t)+\frac{\big(\theta_1-\gamma(1-\delta)\mu_1\big)\Vert\hat{u}[t]\Vert^2}{\gamma\delta m(t)}
        \\&+\frac{\big(\theta_2-\gamma (1-\delta)\mu_2\big)X^2(t)}{\gamma\delta m(t)}+\frac{\big(\theta_3-\gamma(1-\delta)\mu_3\big)\tilde{u}_x^2(s(t),t)}{\gamma m(t)}+\eta\Phi(t)\\&
        -\frac{\big(\mu_1\Vert \hat{u}[t]\Vert^2+\mu_2X^2(t)+\mu_3\tilde{u}_x^2(s(t),t)\big)}{m(t)}\Phi(t).
    \end{split}
\end{equation}
Let us choose $\mu_1,\mu_2,\mu_3$ as in \eqref{betas}, where $\theta_1,\theta_2,\theta_3$ are given by \eqref{thetas}. Also note that the last term in the right hand side of \eqref{ggdum} is negative. Therefore, we have
\begin{equation}\label{ggdum1}
    \begin{split}
        &\dot{\Phi}(t)\leq \frac{(1+\theta_0+\gamma(1-\delta)\sigma) d^2(t)}{\gamma\delta m(t)}+\frac{1-\delta}{\delta}\eta+\frac{\gamma \delta\sigma d^2(t)}{\gamma\delta m(t)}\Phi(t)+\eta\Phi(t),
    \end{split}
\end{equation}
which we rewrite to obtain
\begin{equation}\label{ppq1}
\begin{split}
    \dot{\Phi}(t)\leq& \big(1+\theta_0+\gamma(1-\delta)\sigma\big)\frac{\big(d^2(t)-\gamma(1-\delta)m(t)\big)}{\gamma\delta m(t)}+\big(1+\theta_0+\gamma(1-\delta)\sigma\big)\frac{(1-\delta)}{\delta}\\&+\frac{(1-\delta)\eta}{\delta}+\eta\Phi(t)
    +\gamma\delta\sigma\frac{d^2(t)-\gamma(1-\delta)m(t)}{\gamma\delta m(t)}\Phi(t)+\gamma (1-\delta)\sigma\Phi(t).
\end{split}
\end{equation}
Rearranging the terms in \eqref{ppq1} lead to 
\begin{equation}
    \dot{\Phi}(t)\leq a_1\Phi^2(t)+a_2\Phi(t)+a_3,
\end{equation}
where
\begin{align}
\label{a1}a_{1}&=\gamma\delta\sigma >0,\\
\label{a2}a_{2}&=1+\theta_{0}+2\gamma(1-\delta)\sigma +\eta>0,\\
\label{a3}a_{3}&=\big(1+\theta_{0}+\gamma(1-\delta)\sigma +\eta\big)\frac{1-\delta}{\delta}>0.
\end{align}
By the Comparison principle, it follows that the time needed for $\Phi$ to go from $\Phi(\hat{t}_{j})=0$ to $\Phi(t_{j+1})=1$ is at least 
\begin{equation}\label{mdt}
\tau=\int_{0}^{1}\frac{1}{a_{1}s^{2}+a_{2}s+a_{3}}ds>0.
\end{equation}
Therefore, $t_{j+1}-\hat{t}_{j}\geq \tau$. As $t_{j+1}-t_{j}\geq t_{j+1}-\hat{t}_{j}$, we can conclude that $t_{j+1}-t_{j} \geq \tau$. Note from \eqref{mdt}, \eqref{a3}, and \eqref{delts_cn} that $\tau< 1/a_3< \delta/(1-\delta)<1/c$ where $1/c$ is the maximum dwell-time allowed by the definition of the event-trigger in Definition \ref{def1}. Furthermore, the set of event-times $I$ is a strictly increasing sequence to infinity, \textit{i.e.,} $t_{j}\rightarrow \infty$ as $j\rightarrow \infty$. \hfill $\qed$

\subsubsection{Exponential Convergence}
\begin{thme}\label{kkhr} Consider the closed-loop system containing the plant \eqref{he1},\eqref{he3},\eqref{he4},\eqref{he5},\eqref{lln},\eqref{she} and the observer \eqref{ob1},\eqref{ob3},\eqref{obg},\eqref{lln},\eqref{she1} subject to Assumption \ref{ass1} and conditions \eqref{ooly}-\eqref{cn3}. In Definition \ref{def1}, let us select the event-trigger parameters $\eta,\gamma,\sigma,\mu_1,\mu_2,\mu_3>0$ as follows. Let $\eta,\gamma>0$ be design parameters, and let us choose the parameters $\mu_{1},\mu_{2},\mu_3>0$ as in \eqref{betas}. Further, let us choose 
$\sigma>0$ such that
\begin{equation}\label{fe2}
\sigma = 4A\alpha L,
\end{equation}
where $A$ is any positive parameter that satisfies \begin{equation}\label{fee1}
\begin{split}
A>\max\bigg\{&\frac{96\mu_1L^2}{\alpha}\Big(1+\frac{\beta^2(\zeta^2+\varepsilon^2)L^2}{\alpha^2}\Big),\frac{4\beta\big(3\mu_1(\zeta^2+\varepsilon^2)L+\mu_2\big)}{\varepsilon\alpha c}\bigg\}.
\end{split}
\end{equation}
with $\zeta$ given by \eqref{sko} and $\varepsilon>0$ in \eqref{fdbt2} chosen to satisfy
\begin{equation}\label{pp1}
\varepsilon <\min\bigg\{\frac{\sqrt{\alpha c}}{\beta},\frac{\alpha}{8}\bigg(\beta L\Big(8+\frac{\beta^2R^2L^2}{\alpha^2}\Big)\bigg)^{-1},\varepsilon^*\bigg\},
\end{equation}
where
\begin{equation}\label{pp2}
R=\frac{2\sqrt{\alpha c}}{\beta},
\end{equation}and $\varepsilon^*$ is the positive solution of the quadratic equation
\begin{equation}\label{zzs}
\begin{split}
h(\varepsilon^{*})=&\frac{\alpha c}{4\beta}-\Big(\frac{4\beta^2 R^2L}{\alpha}+\frac{7\alpha}{16L}\Big)\varepsilon^{*}-\Big(4\beta+\frac{\beta^3R^2L^2}{2\alpha^2}\Big)\varepsilon^{*2}=0.
\end{split}
\end{equation}
Then, subject to the event-triggered boundary control apprach in Definition \ref{def1}, the closed-loop system has a unique solution satisfying \eqref{mv1}-\eqref{dots} and exponentially converges to zero, \textit{i.e.,} $\Vert\hat{u}[t]\Vert+\Vert\tilde{u}[t]\Vert+\Vert\tilde{u}_x[t]\Vert+\vert X(t)\vert\rightarrow 0$ as $t\rightarrow \infty$ where $\hat{u}[t]=\hat{T}[t]-T_m, \tilde{u}[t]=T[t]-\hat{T}[t]$, and $X(t)=s(t)-s_r$. 
\end{thme}

\textit{Proof:} Theorem \ref{ffg} ensures the existence of an increasing sequence as it is proven that $t_{j+1}-t_j\geq \tau>0$ for any $j\in\mathbb{N}$. Therefore, considering Lemma \ref{lemcon2}, we can show that the closed-loop system has a unique solution satisfying \eqref{mv1}-\eqref{dots} for all $t>0$.  Now let us show the exponential convergence of the closed-loop system. 

Let us consider a positive definite function $V_1(t)$ involving the system \eqref{what1}-\eqref{whatx} and \eqref{tildw}-\eqref{tildwq} as follows:
\begin{align}\label{ff1}
    V_1 =& \frac{1}{2}\int_{0}^{s(t)}\hat{w}^2(x,t)dx+\frac{\varepsilon\alpha}{2\beta}X^2(t)+\frac{1}{2}\int_{0}^{s(t)}\tilde{w}^2(x,t)dx+\frac{B}{2}\int_{0}^{s(t)}\tilde{w}_x^2(x,t)dx.
\end{align}
where $B>0$. Differentiating \eqref{ff1} along the solution of \eqref{what1}-\eqref{whatx},\eqref{tildw}-\eqref{tildwq} in $t\in[t_j,t_{j+1}),j\in\mathbb{N}$ and using integration by parts and \eqref{intf}, we can obtain that 
\begin{equation}\label{v1dot}
\begin{split}
    \dot{V}_1=&\dot{s}(t)\Big(\frac{\varepsilon^2}{2}X^2(t)+\frac{c}{\beta}X(t)\int_{0}^{s(t)}\hat{w}(x,t)dx\Big)\\&
+\Big(\int_{0}^{s(t)}f(x,s(t))\hat{w}(x,t)dx-\varepsilon\alpha X(t)\Big)\tilde{w}_x(s(t),t)\\&-\alpha\Vert\hat{w}_x[t]\Vert^2-\frac{\varepsilon\alpha c}{\beta}X^2(t)-\varepsilon\beta\psi(-s(t))X(t)\hat{w}(0,t)        \\&-\frac{\varepsilon\beta^2}{\alpha}\int_{x}^{s(t)}\psi(-y)\hat{w}(y,t)dy\hat{w}(0,t)+\varepsilon\beta\hat{w}^2(0,t)\\&-\frac{B}{2}\tilde{w}_x^2(s(t),t)\dot{s}(t)-(\alpha+\lambda B)\Vert\tilde{w}_x[t]\Vert^2-\lambda\Vert\tilde{w}[t]\Vert^2\\&-\alpha B\Vert\tilde{w}_{xx}[t]\Vert^2-\alpha\hat{w}(0,t)d(t).
\end{split}
\end{equation}
From \eqref{psi1}, it follows that $\psi(-x)=e^{-\nu x}(-\zeta\sin(\omega x)+\varepsilon\cos(\omega x)),$ from which we can obtain that $\vert\psi(-x)\vert\leq \sqrt{\zeta^2+\varepsilon^2}.$ Now,  recalling from \eqref{pp1} that $\varepsilon < \frac{\sqrt{\alpha c}}{\beta}$, we can show that $$\zeta^2=\frac{\big(2\alpha c-(\varepsilon\beta)^2\big)^2}{\beta^2\big(4\alpha c-(\varepsilon\beta)^2\big)}< \frac{\big(2\alpha c+\alpha c\big)^2}{(4\alpha c-\alpha c)\beta^2}< \frac{3\alpha c}{\beta^2}.$$ Further, we can write  that $\zeta^2+\varepsilon^2< \frac{3\alpha c}{\beta^2}+\frac{\alpha c}{\beta^2}=\frac{4\alpha c}{\beta^2},$ from which we can obtain that $\vert\psi(-x)\vert< \frac{2\sqrt{\alpha c}}{\beta}=R$. Thus, using Cauchy-Schwarz inequalities on \eqref{v1dot} and noting that $\dot{s}(t)\geq 0$ from Lemma \ref{lemcon2}, we can obtain that 
\begin{equation}\label{jjnk}
\begin{split}
\dot{V}_1\leq &\dot{s}(t)\bigg(\frac{cL\kappa_1}{2\beta}\Vert\hat{w}[t]\Vert^2+\Big(\frac{\varepsilon^2}{2}+\frac{c}{2\beta\kappa_1}\Big)X^2(t)\bigg)\\&+
        \frac{\alpha}{2\kappa_2}\hat{w}^2(0,t)+\frac{\alpha\kappa_2}{2}d^2(t)+\varepsilon\beta\hat{w}^2(0,t)+\frac{\varepsilon\beta R}{2\kappa_3}\hat{w}^2(0,t)\\&+\frac{\varepsilon\beta R\kappa_3}{2}X^2(t)+\frac{\varepsilon\beta^2 R}{2\alpha\kappa_4}\hat{w}^2(0,t)+\frac{\varepsilon\beta^2 RL\kappa_4}{2\alpha}\Vert\hat{w}[t]\Vert^2\\&-\alpha\Vert\hat{w}_x[t]\Vert^2-\frac{\varepsilon\alpha c}{\beta}X^2(t)+\frac{1}{2\kappa_5}\tilde{w}_x^2(s(t),t)\\&+\frac{\kappa_5}{2}f^2_{max}\Vert\hat{w}[t]\Vert^2
    +\frac{\varepsilon\alpha\kappa_6}{2}\tilde{w}_x^2(s(t),t)+\frac{\varepsilon\alpha}{2\kappa_6}X^2(t)\\&-\lambda \Vert\tilde{w}[t]\Vert^2-(\alpha+\lambda B)\Vert \tilde{w}_x[t]\Vert^2-\alpha B\Vert\tilde{w}_{xx}[t]\Vert^2,
        \end{split}
        \end{equation}where $\kappa_1,\kappa_2,\kappa_3,\kappa_4,\kappa_5,\kappa_6$ are some positive constants and $$f_{max}=\sqrt{\max_{0\leq s(t)\leq L}\int_{0}^{s(t)}f^2(x,s(t))dx},$$  for $f(x,s(t))$ given by \eqref{uuj}. Using Poincare and Agmon inequalities together with the fact that $0<s(t)< s_r < L$ for the system \eqref{what1}-\eqref{what3}, we can obtain that
\begin{align}\label{ppjk}
    \Vert\hat{w}[t]\Vert^2&\leq 2L\varepsilon^2 X^2(t)+4L^2\Vert\hat{w}_x[t]\Vert^2,\\\label{ppjk1}
    \hat{w}^2(0,t)&\leq 2\varepsilon^2X^2(t)+4L\Vert \hat{w}_x[t]\Vert^2.
\end{align} 
Substituting $\Vert\hat{w}[t]\Vert^2$ and $\hat{w}^2(0,t)$ in \eqref{jjnk} with \eqref{ppjk} and \eqref{ppjk1}, we can obtain
\begin{equation}\label{aco}
    \begin{split}
       \dot{V}_1\leq &\dot{s}(t)\bigg(\frac{cL\kappa_1}{2\beta}\Vert\hat{w}[t]\Vert^2+\Big(\frac{\varepsilon^2}{2}+\frac{c}{2\beta\kappa_1}\Big)X^2(t)\bigg)+\frac{\alpha\kappa_2}{2} d^2(t)\\ &
        -\varepsilon \Big(\frac{\alpha c}{\beta}-\frac{\beta R\kappa_3}{2}-\frac{\alpha}{2\kappa_6}-\frac{\varepsilon^2\beta^2RL^2\kappa_4}{\alpha}-L\varepsilon\kappa_5f_{max}^2\\&\qquad\qquad\qquad\qquad-\frac{\varepsilon\alpha}{\kappa_2}-2\varepsilon^2\beta-\frac{\varepsilon^2\beta R}{\kappa_3}-\frac{\varepsilon^2\beta^2R}{\alpha\kappa_4}\Big)X^2(t)\\&
        -\Big(\alpha-\frac{2\varepsilon\beta^2 RL^3\kappa_4}{\alpha}-2L^2\kappa_5f_{max}^2-\frac{2\alpha L}{\kappa_2}-4L\varepsilon\beta-\frac{2L\varepsilon\beta R}{\kappa_3}-\frac{2L\varepsilon\beta^2R}{\alpha\kappa_4}\Big)\Vert\hat{w}_x[t]\Vert^2\\&+\Big(\frac{4L^2f_{max}^2}{\alpha}+\frac{\varepsilon\alpha\beta}{2c}\Big)\tilde{w}_x^2(s(t),t)-\lambda \Vert\tilde{w}[t]\Vert^2-(\alpha+\lambda B)\Vert \tilde{w}_x[t]\Vert^2-\alpha B\Vert\tilde{w}_{xx}[t]\Vert^2.
    \end{split}
\end{equation}
Let us choose $\kappa_1=1,\kappa_2=8L,\kappa_3=\frac{8\varepsilon\beta RL}{\alpha},\kappa_4=\frac{\beta R}{2\alpha},\kappa_5=\frac{\alpha}{8L^2f_{max}^2},\kappa_6=\frac{\beta}{c}$. Then, we can rewrite \eqref{aco} as
\begin{equation}\label{taca}
    \begin{split}
       \dot{V}_1\leq &\dot{s}(t)\bigg(\frac{cL}{2\beta}\Vert\hat{w}[t]\Vert^2+\Big(\frac{\varepsilon^2}{2}+\frac{c}{2\beta}\Big)X^2(t)\bigg)+4\alpha Ld^2(t)
       \\&-\varepsilon \Big(\frac{\alpha c}{2\beta}-\frac{4\varepsilon\beta^2R^2L}{\alpha}-\frac{\varepsilon^2\beta^3R^2L^2}{2\alpha^2}-\frac{3\varepsilon\alpha}{8L}-4\varepsilon^2\beta\Big)X^2(t)\\&
       -\bigg(\frac{\alpha}{4}-\varepsilon\beta L\Big(8+\frac{\beta^2R^2L^2}{\alpha^2}\Big)\bigg)\Vert\hat{w}_x[t]\Vert^2
       \\&  
       +\Big(\frac{4L^2f_{max}^2}{\alpha}+\frac{\varepsilon\alpha\beta}{2c}\Big)\tilde{w}_x^2(s(t),t)-\lambda \Vert\tilde{w}[t]\Vert^2-(\alpha+\lambda B)\Vert \tilde{w}_x[t]\Vert^2-\alpha B\Vert\tilde{w}_{xx}[t]\Vert^2.
    \end{split}
\end{equation}

Then, recalling from \eqref{pp1} that $\frac{\alpha}{8}>\varepsilon\beta L\Big(8+\frac{\beta^2R^2L^2}{\alpha^2}\Big),$ letting $B=\frac{4L^2 f_{max}^2}{\alpha^2}+\frac{\varepsilon\beta}{2c}+b^{*}$ where $b^*>0$, and substituting $\Vert \hat{w}_x[t]\Vert^2$ with \eqref{ppjk}, we can obtain from \eqref{taca} that 
\begin{equation}\label{pldy}
\begin{split}
\dot{V}_1\leq &\dot{s}(t)\bigg(\frac{cL}{2\beta}\Vert\hat{w}[t]\Vert^2+\Big(\frac{\varepsilon^2}{2}+\frac{c}{2\beta}\Big)X^2(t)\bigg)+4\alpha Ld^2(t)\\&
         -\varepsilon\Big(\frac{\alpha c}{2\beta}-\frac{4\varepsilon\beta^2R^2 L}{\alpha}-\frac{7\varepsilon\alpha}{16L}-4\varepsilon^2\beta-\frac{\varepsilon^2\beta^3R^2L^2}{2\alpha^2}\Big)X^2(t)\\&-\frac{\alpha}{32L^2}\Vert\hat{w}[t]\Vert^2-\alpha b^*\Vert\tilde{w}_{xx}[t]\Vert^2-\lambda\Vert\tilde{w}[t]\Vert^2-(\alpha+\lambda B)\Vert\tilde{w}_x[t]\Vert^2.
         \end{split}
         \end{equation}
Above we have used the fact that $\tilde{w}_x^2(s(t),t)\leq \Vert\tilde{w}_{xx}[t]\Vert^2$. Let us define $$h(\varepsilon)=\frac{\alpha c}{4\beta}-\Big(\frac{4\beta^2 R^2L}{\alpha}+\frac{7\alpha}{16L}\Big)\varepsilon-\Big(4\beta+\frac{\beta^3R^2L^2}{2\alpha^2}\Big)\varepsilon^{2}.$$ 
Then, we have that $ h(0)=\frac{\alpha c}{4\beta}>0$ and 

$$h'(\varepsilon)=-\frac{4\beta^2R^2 L}{\alpha}-\frac{7\alpha}{16L}-8\beta\varepsilon-\frac{\beta^3 R^2L^2}{\alpha^2}\varepsilon<0$$ for $\varepsilon\geq 0.$ Therefore, we have $\varepsilon^*$ such that $h(\varepsilon)>0$ for $0\leq \varepsilon\leq \varepsilon^*$ and $h(\varepsilon^*)=0$. Since $\varepsilon<\varepsilon^*$ (see \eqref{pp1}), we have that $h(\varepsilon)>0$. Thus, we can write from \eqref{pldy} that 
\begin{equation}\label{dotv33}
\begin{split}
        \dot{V}_1\leq &\dot{s}(t)\bigg(\frac{cL}{2\beta}\Vert\hat{w}[t]\Vert^2+\Big(\frac{\varepsilon^2}{2}+\frac{c}{2\beta}\Big)X^2(t)\bigg)+4\alpha Ld^2(t)\\&
         -\varepsilon\Big(\frac{\alpha c}{4\beta}+h(\varepsilon)\Big)X^2(t)-\frac{\alpha}{32L^2}\Vert\hat{w}[t]\Vert^2-\alpha b^*\Vert\tilde{w}_{xx}[t]\Vert^2\\&-\lambda\Vert\tilde{w}[t]\Vert^2-(\alpha+\lambda B)\Vert\tilde{w}_x[t]\Vert^2.
\end{split}
\end{equation}
Let us consider the following Lyapunov function recalling from Lemma \ref{pom} that $m(t)>0$ for $t\geq 0$:
\begin{equation}\label{tdf}
    V=AV_1+m.
\end{equation}
Taking the time derivative of \eqref{tdf} and using \eqref{dotv33}, we can write that 
\begin{equation}\label{yyu}
\begin{split}
  \dot{V}\leq & A\dot{s}(t)\bigg(\frac{cL}{2\beta}\Vert\hat{w}[t]\Vert^2+\Big(\frac{\varepsilon^2}{2}+\frac{c}{2\beta}\Big)X^2(t)\bigg)+4A\alpha Ld^2(t)\\&-A\varepsilon\Big(\frac{\alpha c}{4\beta}+h(\varepsilon)\Big)X^2(t)-\frac{A\alpha}{32L^2}\Vert\hat{w}[t]\Vert^2-A\alpha b^*\Vert\tilde{w}_{xx}[t]\Vert^2\\&-A\lambda\Vert\tilde{w}[t]\Vert^2-A(\alpha+\lambda B)\Vert\tilde{w}_x[t]\Vert^2+\dot{m}(t).
  \end{split}
  \end{equation}
  Noting that $\dot{s}(t)\geq 0$ from Lemma \ref{lemcon2}, using \eqref{ff1} along with the dynamics of $m(t)$ given by \eqref{obetbc32}, we can obtain from \eqref{yyu} that
\begin{equation}\label{muriels}
  \begin{split}
  \dot{V}\leq &\xi\dot{s}(t)V-A\varepsilon\Big(\frac{\alpha c}{4\beta}+h(\varepsilon)\Big)X^2(t)-\frac{A\alpha}{32L^2}\Vert\hat{w}[t]\Vert^2\\&-A\alpha b^*\Vert\tilde{w}_{xx}[t]\Vert^2-A\lambda\Vert\tilde{w}[t]\Vert^2-A(\alpha+\lambda B)\Vert\tilde{w}_x[t]\Vert^2\\&+(4A\alpha L-\sigma)d^2(t)-\eta m(t)+\mu_1\Vert\hat{u}[t]\Vert^2+\mu_2X^2(t)+\mu_3\tilde{u}_x^2(s(t),t),
  \end{split}
  \end{equation}
  where
  \begin{equation}
      \xi=\max\bigg\{\frac{cL}{\beta},\frac{\beta}{\alpha\varepsilon}\Big(\varepsilon^2+\frac{c}{\beta}\Big)\bigg\}.
  \end{equation}
  Using Cauchy-Schwarz inequality and Young's inequality along with the fact that $0<x<s(t)<L$, we can obtain from \eqref{intf} that 
  \begin{equation}
  \begin{split}
  &\Vert\hat{u}[t]\Vert^2\leq 3\Vert\hat{w}[t]\Vert^2+\frac{3\beta^2(\zeta^2+\varepsilon^2)L^2}{\alpha^2}\Vert \hat{w}[t]\Vert^2+3(\zeta^2+\varepsilon^2)LX^2(t).
  \end{split}
  \end{equation}
  Noting that $\tilde{u}_x(s(t),t)=\tilde{w}_x(s(t),t)$ and that $\tilde{w}_x^2(s(t),t)\leq \Vert\tilde{w}_{xx}[t]\Vert^2$, we can obtain from \eqref{muriels} that 
  \begin{equation}\label{kkhj}
  \begin{split}
  \dot{V}\leq &\xi\dot{s}(t) V-\bigg(A\varepsilon\Big(\frac{\alpha c}{4\beta}+h(\varepsilon)\Big)-3\mu_1(\zeta^2+\varepsilon^2)L-\mu_2\bigg)X^2(t)\\&
    -\bigg(\frac{A\alpha}{32L^2}-3\mu_1\Big(1+\frac{\beta^2(\zeta^2+\varepsilon^2)L^2}{\alpha^2}\Big)\bigg)\Vert\hat{w}[t]\Vert^2\\&-(A\alpha b^*-\mu_3)\Vert \tilde{w}_{xx}[t]\Vert^2-A\lambda \Vert\tilde{w}[t]\Vert^2\\&-A(\alpha+\lambda B)\Vert\tilde{w}_x[t]\Vert^2-\eta m(t)+(4A\alpha L-\sigma)d^2(t).
\end{split}
\end{equation}
Let us define 
\begin{equation}
    b_{1}:=\frac{A\alpha}{32L^2}-3\mu_1\Big(1+\frac{\beta^2(\zeta^2+\varepsilon^2)L^2}{\alpha^2}\Big),
\end{equation}
and 
\begin{equation}
    b_{2}:=A\varepsilon\Big(\frac{\alpha c}{4\beta}+h(\varepsilon)\Big)-3\mu_1(\zeta^2+\varepsilon^2)L-\mu_2.
\end{equation}
Recall that $h(\varepsilon)>0$ due to the choice of $\varepsilon$ as in \eqref{pp1}-\eqref{zzs}. Considering \eqref{fee1}, we can show that $b_{1},b_{2}>0$. Let us select $b^*>0$ such that $b^*>\frac{\mu_3}{A\alpha}.$ Then, recalling \eqref{fe2}, we can write from \eqref{kkhj} that
\begin{equation}\label{oops}
    \dot{V}\leq \xi\dot{s}(t)V-2b V, \text{ }t\in(t_j,t_{j+1}),
\end{equation}
where 
\begin{equation}
    b=\min\Big\{\frac{b_{1}}{A},\frac{b_{2}\beta}{A\varepsilon\alpha},\lambda,\frac{\eta}{2}\Big\}.
\end{equation}
Consider the following functional
\begin{equation}\label{nmw11}
W=Ve^{-\xi s(t)}.
\end{equation}
Taking the time derivative of \eqref{nmw11} for $t\in(t_{j},t_{j+1})$ with the aid of \eqref{oops}, we can deduce $\dot{W}\leq -2bW.$ Via integration and considering \eqref{nmw11} and the fact that $0<s_0\leq s(t)<L$, we can obtain $$V(t)\leq e^{-2bt}e^{\xi \big(s(t)-s_0\big)}V_s(0)\leq e^{-2bt}e^{\xi L}V_s(0).$$ Thus, using the transformations \eqref{obeft},\eqref{inobst},\eqref{fdbt1},\eqref{intf}, Young's and Cauchy-Schwarz inequalities, we can show that there exists a constant $M>0$ such that \begin{align}
    \Vert\hat{u}[t]\Vert+\Vert\tilde{u}[t]\Vert&+\Vert\tilde{u}_x[t]\Vert+\vert X(t)\vert\\&
    \leq Me^{-bt}
   \sqrt{\big(\Vert\hat{u}[0]\Vert^2+\Vert\tilde{u}[0]\Vert^2+\Vert\tilde{u}_x[0]\Vert^2+\vert X(0)\vert^2+m(0)\big)}.
\end{align} 
\hfill $\qed$

\begin{rmk}\label{rmkn}\textit{(Selection of the event-trigger parameters $\eta,\gamma,\sigma,\mu_1,\mu_2,\mu_3$)} The parameters $\mu_1,\mu_2,\mu_3>0$ are chosen to satisfy \eqref{betas} with $\gamma>0$ being treated as a free parameter which can be chosen to scale up/down the values of $\mu_1,\mu_2,\mu_3$ as required, and $\delta\in(0,1)$ chosen to satisfy \eqref{delts_cn}. The parameter $\sigma>0$ is chosen as in \eqref{fe2}. The parameter $\eta>0$ is also a free parameter, which can be used to adjust the convergence rate of the dynamic variable $m(t)$.  
\end{rmk}

\section{Numerical Simulations and Discussion}

We carry out simulations for the one-phase Stefan problem considering a cylinder of paraffin with length $L=3.0\text{ }$cm whose physical parameters are as follows: $\rho=790\text{ }$kg.m$^{-3};{ }$$\Delta H^*=210\text{ }$J.g$^{-1};\text{ }$$C_p=2.38\text{ }$J.g$^{-1}. ^\circ$C$^{-1};\text{ }T_m=37.0\text{ }^\circ$C$;\text{ }$$k=0.220\text{ }$W.m$^{-1}$. We use a semi-implicit numerical scheme relying on the so-called boundary immobilization method \citep{koleva2010numerical}. A uniform step size of $h=0.05$ is used for the space variable and a uniform step size of $0.5$ s is used for the time variable. The setpoint and the initial values are chosen as $s_r=2.0$ cm, $s_0=0.1$ cm, $T_0(x)-T_m=(1-x/s_0)$ and $\hat{T}_0(x)-T_m=10(1-x/s_0)$. The observer gain $\lambda$ is chosen as $\lambda = 0.1$. Note that under these choices of $s_r,s_0,T_0(x),\hat{T}_0(x),$ and $\lambda$, there exists constants $\hat{H}_u\geq\hat{H}_\ell>H$ such that the conditions \eqref{ooly}-\eqref{cn3} are satisfied. The control gain in \eqref{lln} is chosen as $c=3.0\times 10^{-4}/s$. The parameter $\varepsilon$ in \eqref{fdbt2} is chosen as $\varepsilon=10$ such that \eqref{pp1}-\eqref{zzs} are satisfied. The parameters for the event-trigger are chosen as follows: $\eta = 1.325\times 10^{-2};\gamma =10^3; \mu_1=1.42\times 10^{-4};\mu_2=36.85;\mu_3=2.2079\times 10^{14};\sigma =6.19\times 10^{-5}; m(0)=10^{-4}$. Note that the above choices of $\mu_1,\mu_2,\mu_3$ satisfy \eqref{betas} when $\delta=0.5$. The chosen $\sigma$ satisfies \eqref{fe2} when $A=4.42\times 10^3$.

\begin{figure}
\centering
\includegraphics[scale=0.5]{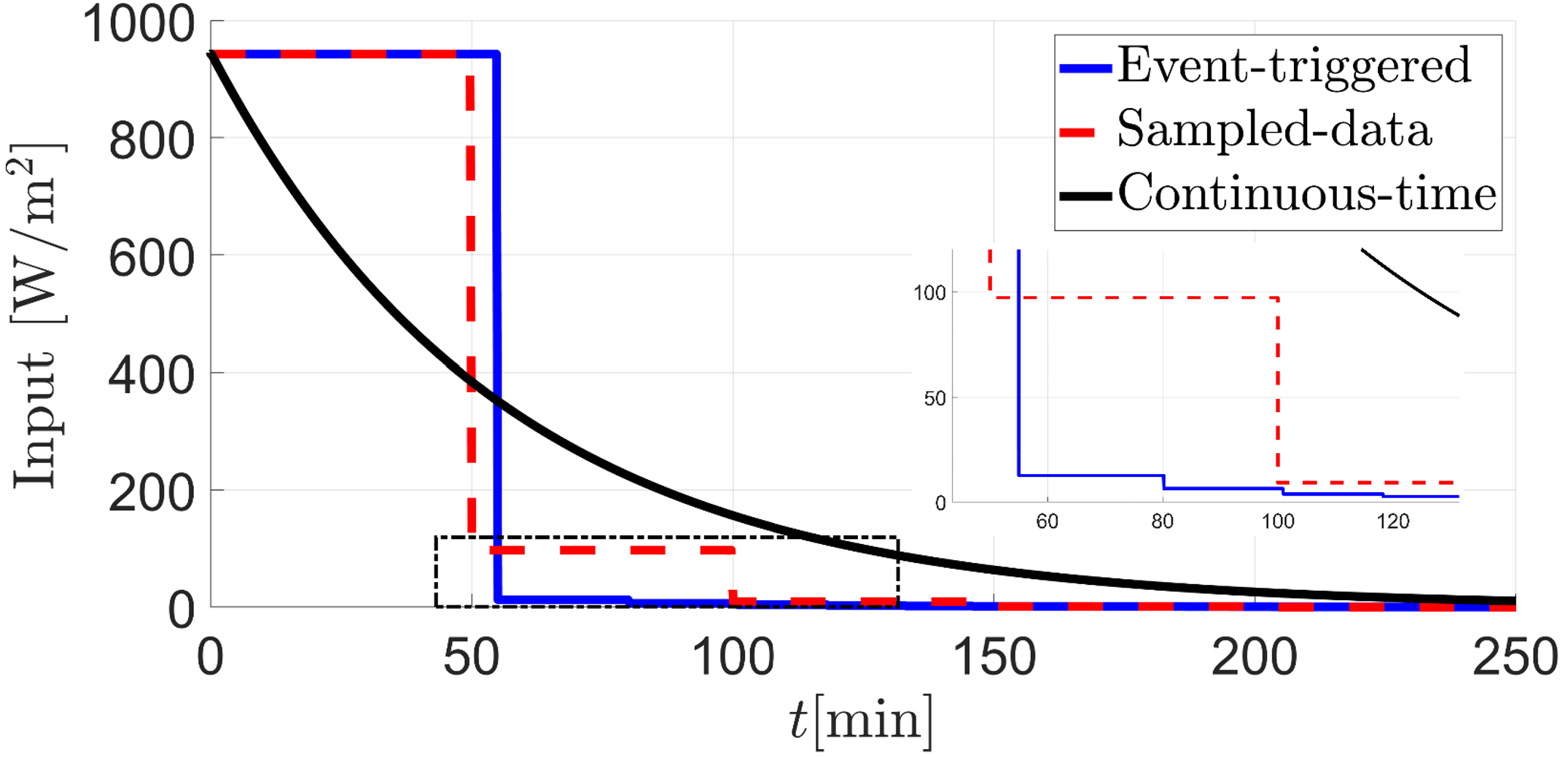}
\caption{Control inputs under event-triggered control, sampled-data control, and continuous-time control. Sampled-data control with sampling period of $50$[min] is shown for demonstration purpose only since its convergence guarantees under observer-based setting has not been established.} 
\label{figuu}
\end{figure}

\begin{figure}
\centering
\includegraphics[scale=0.63]{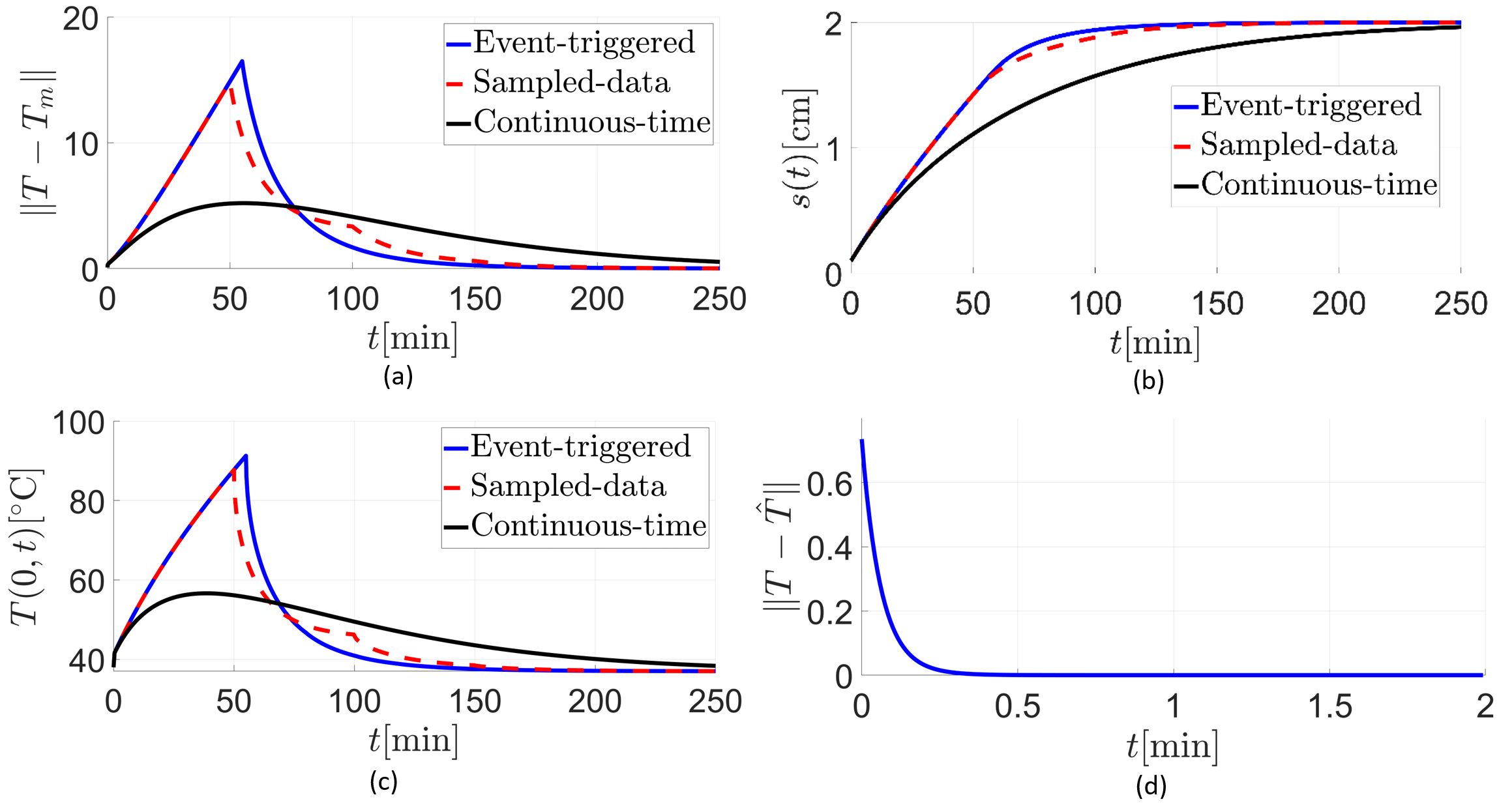}
\caption{Closed-loop system signals (a) $\Vert T-T_m\Vert$, (b) interface position $s(t)$[cm], (c) $T(0,t)[^\circ C]$, (d) $\Vert T-\hat{T}\Vert$.} 
\label{fig1w}
\end{figure}

The control inputs under even-triggered, sampled-data, and continuous-time control are shown in Fig. \ref{figuu}. 
We can observe that event-triggered control eliminates the necessity
of continuous control updates. Fig. \ref{fig1w} illustrates the time evolution of various closed-loop system signals under the proposed event-based boundary control. The time responses of $\Vert T-T_m\Vert$, boundary temperature $T(0,t)$[$^\circ$C], and $\Vert T-\hat{T}\Vert$ are shown in Fig. 4(a), Fig. 4(c), and Fig. 4(d), respectively. It can be observed that, $\Vert T-T_m\Vert\rightarrow 0$ and $\Vert T-\hat{T}\Vert\rightarrow 0$ as $t\rightarrow\infty$ and $T(0,t)\geq T_m$ for all $t\geq 0$. Fig. 4(b) shows the the time evolution of the interface position under the proposed event-based boundary control. The interface position $s(t)$ converges to the setpoint $s_r$ monotonically without overshooting (which also confirms that $\dot{s}(t)\geq 0$). 

\section{Conclusion}
In this paper, we have proposed an observer-based event-triggered boundary control strategy for the one-phase Stefan problem using the infinite-dimensional backstepping approach. We have used a dynamic event-triggering condition to determine the time instances when the control input needs to be updated. Under the observer-based event-triggered control strategy, we have proved the existence of a uniform minimal dwell-time, which excludes Zeno behavior from the closed-loop system. Furthermore, we have proved the well-posedness of the closed-loop system along with the model validity conditions. Finally, we have shown that the proposed control approach exponentially converges the closed-loop system to the setpoint. In our future work, we will consider the observer-based event-triggered boundary control of the two-phase Stefan problem. 

\section*{Disclosure statement}
The authors hereby declare that they have no relevant financial or non-financial competing interests in relation to this manuscript submitted to the journal.

\bibliographystyle{apacite}
\bibliography{main}

\end{document}